\documentstyle[12pt]{article}
\def\beq{\begin{equation}}
\def\eeq{\end{equation}}
\def\be{\begin{equation}}
\def\ee{\end{equation}}
\def\bea{\begin{eqnarray}}
\def\eea{\end{eqnarray}}
\def\p{\partial}
\def\n{\nabla}
\def\d{\displaystyle}

\begin{document}

\title{\bf Equations of Motion for Massive Spin 2 Field Coupled to
Gravity}

\author{\sc I.L. Buchbinder${}^{a,b}$, D.M. Gitman${}^a$,
V.A. Krykhtin${}^c$ \\ \sc and V.D. Pershin${}^d$
\footnote{e-mail: \tt ilb@mail.tomsknet.ru, gitman@fma.if.usp.br,
krykhtin@phys.dfe.tpu.edu.ru, pershin@ic.tsu.ru}
\\
\it ${}^a$Instituto de F\'\i sica, Universidade de S\~ao Paulo,\\
\it P.O. Box 66318, 05315-970, S\~ao Paulo, SP, Brasil\\
\it ${}^b$Department of Theoretical Physics,\\
\it Tomsk State Pedagogical University,\\
\it Tomsk 634041, Russia\\
\it ${}^c$Department of Theoretical and Experimental Physics, \\
\it Tomsk Polytechnical University,\\
\it Tomsk 634050, Russia \\
\it ${}^d$Department of Theoretical Physics, Tomsk State
University,\\ \it Tomsk 634050, Russia }

\date{}

\begin{titlepage}

\maketitle
\thispagestyle{empty}

%\vspace{0.5cm}

\begin{abstract}
We investigate the problems of consistency and causality for the
equations of motion describing massive spin two field in external
gravitational and massless scalar dilaton fields in arbitrary
spacetime dimension.  From the field theoretical point of view we
consider a general classical action with non-minimal couplings and
find gravitational and dilaton background on which this action
describes a theory consistent with the flat space limit. In the
case of pure gravitational background all field components propagate
causally. We show also that the massive spin two field can be
consistently described in arbitrary background by means of the
lagrangian representing an infinite series in the inverse mass.
Within string theory we obtain equations of motion for the massive
spin two field coupled to gravity from the requirement of quantum
Weyl invariance of the corresponding two dimensional sigma-model. In
the lowest order in $\alpha'$ we demonstrate that these effective
equations of motion coincide with consistent equations derived in
field theory.
\end{abstract}

{\small

}

\end{titlepage}

\section{Introduction}

The purpose of this paper is twofold - first, to describe consistent
theory of massive spin 2 field interacting with external gravity from
the point of view of classical field theory and, second, to derive
effective equations of motion for this field from string theory.

The problem of consistent description of higher spin fields
interaction has a long history but is still far from the complete
resolution. Within framework of standard field theory
it is possible to construct a classical action for the
higher spin fields only on specific curved spacetime manifolds
\cite{aragone}--\cite{klish2}\footnote{For description of
higher spin massless fields on specific background see e.g.
\cite{vasilev,vasilev2}, the case of a collection of massless spin-2
fields was investigated in \cite{wald}}.  For example, massive
integer spins are described by symmetric tensors of corresponding
ranks and one can write down the equations of motion and find
classical actions for them but only in special spacetimes (e.g. in
Ricci flat spaces).  It means that in such an appoach gravity field
does not feel the presence of higher spins matter through an
energy-momentum tensor. So a consistent classical action for the
system of dynamical gravity and a higher massive field is still
unknown and there are some indications that perhaps it does not exist
at all.

One of these indications comes from considering a Kaluza-Klein
decomposition of Einstein gravity in $D-$dimensional
spacetime into gravity plus infinite tower of massive second rank
tensor fields in $(D-1)-$dimensional world, masses being
proportional to inverse compactification radius. Then the resulting
four dimensional theory of spin 2 fields interacting with gravity and
with each other should be consistent as one started from the ordinary
Einstein theory and just considers it on a specific manifold.  But as
was shown in \cite{duff}, it is impossible to reduce this theory
consistently to a finite number of spin 2 fields, i.e. consistency can
be achieved only if the whole infinite tower of higher massive
fields are present in the theory.

The main problem in higher spin fields theories is that
introduction of interaction may excite new unphysical degrees of
freedom which lead in general to appearance of negative norm states
in the hilbert space and to violation of causality. In free theories
these unphysical degrees are absent due to transversality and
tracelessness conditions. Lagrangian description of these conditions
requires auxiliary fields vanishing on equations of motion
\cite{fierz,singh}. In arbitrary interacting theories the
auxiliary fields may become dynamical and to make them vanishing
again one has to impose some additional restrictions on the kind of
interaction.

As a first step towards a full dynamical theory of interacting higher
spin fields one usually tries to describe a single higher spin
massive field in an external electromagnetic or gravitational
background \cite{bengt,ovrut,klish2,zwanziger,zwanziger2,3/2}. In
this case one way to achieve consistency is to impose appropriate
restrictions on the external background field.  For example, it is
well known that consistent theories of higher massive spin fields can
be easily built in the spacetimes of constant curvature. In the
Section~2 we describe in detail the theory of massive spin 2 field in
curved spacetime and show how these restrictions on the external
gravitational background arise. As a result we will arrive to a
one-parameter family of lagrangian theories which describe consistent
propagation of the spin 2 field in arbitrary Einstein spacetime.

There exists another possible way to achieve consistency by
constructing equations for higher massive fields in form of infinite
series in inverse mass (or, equivalently, in curvature). A recent
attempt in this direction was undertaken in \cite{klish2} where,
however, only theories on symmetric Einstein manifolds were
considered and consistent equations were derived in the simplest
approximation linear in curvature.

In this paper we demonstrate that consistent equations for the spin 2
massive field can, in principle, be constructed as infinite series in
inverse mass square in arbitrary gravitational background. These kinds
of infinite series arise naturally in string theory which represent
another approach for consistent description of higher spins
interaction.

String theory contains an infinite number of massive
fields with various spins interacting with each other and with a
finite number of massless fields.  Unfortunately, there are arguments
that in string theory a general coordinate invariant effective field
action reproducing the correct S-matrix both for massless and massive
string states does not exist \cite{versus}. The full effective action
for all string fields is not general coordinate invariant and general
covariance arises only as an approximate symmetry in effective action
for massless fields once all the massive fields are integrated out.
That effective action for massive fields cannot be covariant follows,
for example, from the fact that terms cubic in massive fields can
contain only flat metric and there is no terms of higher powers
\cite{versus}.

Influence of massive string modes is negligible at low
energies but they become important, for instance, in string cosmology
\cite{maggiore} (for a recent review see e.g. \cite{veneziano}) so it
would be desirable to have consistent equations describing
interaction of the massive string fields with gravity.

And indeed, there exists a possibility to derive from the string
theory some covariant parts of equations for massive higher spins
fields interacting with background gravity. The aim of our paper
is to show explicitly how this procedure works using as an example
dynamics of the second rank tensor from the first massive level of
open bosonic string.

A convenient method of deriving effective field equations of motion
from the string theory is provided by the $\sigma-$model approach
\cite{lovelace}--\cite{curci}. Within this approach a string
interacting with background fields is described by a two dimensional
field theory and effective equations of motion arise from the
requirement of quantum Weyl invariance.  Perturbative derivation of
these equations is well suited for massless string modes because the
corresponding two dimensional theory is renormalizable and loop
expansion corresponds to expansion of string effective action in
powers of string length $\sqrt{\alpha'}$.

Inclusion of interaction with massive modes
\cite{labastida}--\cite{porrati} makes the theory
non-re\-nor\-ma\-li\-zab\-le but this fact does not represent a
problem since in string theory one considers the whole infinite
set of massive fields. Infinite number of counterterms needed for
cancellation of divergences generating by a specific massive field in
classical action leads to renormalization of an infinite number of
massive fields. The only property of the theory crucial for
possibility of derivation of perturbative information is that number
of massive fields giving contributions to renormalization of the
given field should be finite.  As was shown in \cite{bflp} string
theory does fulfill this requirement. To calculate $\beta-$function
for any massive field it is sufficient to find divergences coming
only from a finite number of other massive fields and so it is
possible to derive effective equations of motion for any background
fields in any order in $\alpha'$.

This procedure have serious limitation -- it does not allow to derive
non-perturbative contributions to the string effective action. It is
well known that correct terms cubic in massive fields (quadratic
terms in equations of motion) cannot be found within perturbative
renormalization of $\sigma-$model. For example, perturbative
$\beta-$function of tachyon field is linear in all orders in
$\alpha'$ and interacting tachyonic terms are due to non-perturbative
(from the point of view of two dimensional field theory) effects
\cite{tachyon}.

To find these non-perturbative contribution one should use the method
exact renormalization group (ERG) \cite{banks}--\cite{ellw}. It is
ideally suited for string theory because, first of all, its
formulation does not require any {\em a priori} defined perturbative
scheme, and secondly, the equations of exact renormalization group
are quadratic in interaction terms thus resembling the cubic
structure of exact string field action. But ERG method also has a
serious disadvantage as it requires explicit separation of classical
action into free and interaction parts and thus leads to
non-covariant equations for background fields.  From the general
point of view this fact does not contradict general properties of
string theory. We have already mentioned that exact effective action
describing both massless and massive string modes {\em should} be
non-covariant. But non-covariance of the ERG method makes it rather
difficult to establish relations between string fields equations and
ordinary field theory.

So in this paper we derive covariant equations of motion for massive
string fields interacting with gravity by means of ordinary
perturbative analysis of quantum Weyl invariance condition in the
corresponding $\sigma-$model. Of course, perturbatively we can
obtain equations only linear in massive fields. It was noted long
ago \cite{versus} that one can make such a field redefinition in
the string effective action that terms quadratic in massive fields
(linear terms in equations of motion) acquire dependence on arbitrary
higher powers of massless fields and so may be covariant. In this
paper we explicitly obtain these interaction terms in the
lowest in $\alpha'$ approximation. We do not get terms
quadratic in massive fields which should be non-covariant and arise
only non-perturbatively.

As a model for our calculations we use bosonic open string theory
interacting with background fields of the massless and the first
massive levels. First massive level contains symmetric second
rank tensor and so this model provides the simplest example of
massive higher spin field interacting with gravity. In order to
obtain equations of motion for string fields we build
effective action for the corresponding two dimensional theory,
perform renormalization of background fields and composite operators
and construct the renormalized operator of energy momentum tensor
trace.

This rather standard scheme was first developed for calculations in
closed string theory with massless background fields \cite{cfmp,
tsNPB87, osborn} and then was generalized for the open string theory
\cite{dorn, acny, clny, behrndt} and for strings in massive fields
\cite{bflp}. The new feature appearing in open string theory is that
the corresponding two dimensional sigma model represents a quantum
field theory on a manifold with a boundary. To construct quantum
effective action in such a theory we use generalization of
Schwinger-De~Witt method for manifolds with boundaries developed in
the series of papers \cite{mcavity}.

Making perturbative calculations we restrict ourselves to string
world sheets with topology of a disk. The resulting equations of
motion for graviton will not contain dependence on massive fields
from the open string spectrum because these fields interact only with
the boundary of world sheet and so can not influence the local
physics in the bulk. For example, in the case of graviton and
massive fields from the open string spectrum one expects that
equations of motion for the graviton should look like ordinary vacuum
Einstein equations without any matter. Of course, one can obtain
contributions from open string background fields to the right hand
side of Einstein equations for gravity but it would demand
considering of world sheets of higher genus \cite{fs,clny}.

The organization of the paper is as follows. In the Sec.~2 we
describe the most general consistent equations of motion for massive
spin 2 field on a specific class of curved manifolds from the point
of view of ordinary field theory and generalize them for the
interaction with scalar dilaton field in the Sec.~3. In the next
section we show how the scheme can be generalized to the case of
arbitrary gravitational background by means of either non-lagrangian
equations or equations representing infinite series in inverse mass.
Sec.~5 contains description of the string model that
we use for derivations of effective equations of motion of string
massive fields. We calculate divergences of the theory in the lowest
order, carry out renormalization of background fields and
composite operators and construct the renormalized operator of the
energy-momentum tensor trace. Requirement of quantum Weyl invariance
leads to effective string fields equations of motion. We
compare with each other equations of motion derived in two previous
sections and show that string theory gives (at least in the lowest
order) consistent equations for the spin 2 massive field interacting
with gravity. Conclusion contains summary of the results.

\section{Massive spin 2 field coupled to gravity in field theory}

In this section we give detailed analysis  of the lagrangian
formulation for the free spin 2 massive field and then generalize it
for the presence of external gravitational field. The first
requirement one should impose on such a theory is preservation of the
same number of degrees of freedom and constraints as in the flat
theory. Another important feature of any consistent relativistic
theory is causality, which means that the equations of motion should
not describe superluminal propagation \cite{zwanziger} (see also
\cite{zwanziger2} for a review). The result obtained in this
section is a lagrangian which depends on one arbitrary
dimensionless parameter of non-minimal coupling and describes
consistent propagation of the spin 2 massive field in arbitrary
Einstein spacetime, i.e. it contains the same number of lagrangian
constraints as in the flat spacetime and does not violate causality.

In the flat spacetime the massive spin 2 field is described (as
follows from the analysis of irreducible representations of
4-dimensional Poincare group) by symmetric transversal and traceless
tensor of the second rank $H_{\mu\nu}$ satisfying mass-shell
condition:
\begin{equation}
\Bigr(\partial^2-m^2\Bigl) H_{\mu\nu}=0 {,}\qquad
\partial^\mu H_{\mu\nu}=0 {,}\qquad
H^\mu{}_\mu=0 {.}
\label{irred}
\end{equation}
In higher dimensional spacetimes Poincare algebras have more than two
Casimir operators and so there are several different spins for $D>4$.
Talking about spin 2 massive field in arbitrary dimension we will
mean, as usual, that this field by definition satisfies the same
equations (\ref{irred}) as in $D=4$. After dimensional reduction to
$D=4$ such a field will describe massive spin two representation of
$D=4$ Poincare algebra plus infinite tower of Kaluza-Klein
descendants.

The most convinient approach to building interacting field theories
is a lagrangian one and, in fact, the general point of view is that
any consistent equations of motion should follow from some classical
action. In the case of the free massive spin 2 field it is well known
that all the equations (\ref{irred}) can be derived from the
Fierz-Pauli action \cite{fierz}:
\begin{eqnarray}
S&=&\int\! d^D x \biggl\{ \frac{1}{4} \partial_\mu H \partial^\mu H
-\frac{1}{4} \partial_\mu H_{\nu\rho} \partial^\mu H^{\nu\rho}
-\frac{1}{2} \partial^\mu H_{\mu\nu} \partial^\nu H
+\frac{1}{2} \partial_\mu H_{\nu\rho} \partial^\rho H^{\nu\mu}
\nonumber\\&&
\qquad\qquad
{} - \frac{m^2}{4} H_{\mu\nu} H^{\mu\nu} + \frac{m^2}{4}  H^2
 \biggr\}
\label{actfield}
\end{eqnarray}
where $H=\eta^{\mu\nu} H_{\mu\nu}$.

The general scheme of calculating the complete set of
constraints in lagrangian formalism \cite{gitman} is
equivalent to the Dirac-Bergmann procedure in hamiltonian formalism
and in the case of second class constraints (which is relevant
for massive higher spin fields) consists in the following steps.
If in a theory of some set of fields $\phi^A(x)$, $A=1,\ldots,N$ the
original lagrangian equations of motion define only $r<N$ of
the second time derivaties (``accelerations'') $\ddot\phi^A$ then one
can build $N-r$ primary constraints, i.e. linear combinations of the
equations of motion that does not contain accelerations. Requirement
of conservation in time of the primary constraints either define some
of the missing accelerations or lead to new (secondary) constraints.
Then one demands conservation of the secondary constraints and so on,
until all the accelerations are defined and the procedure closes up.

Applying this procedure to the action (\ref{actfield}) one can see
that the equations of motion
\begin{eqnarray}
E_{\mu\nu}&=&\partial^2 H_{\mu\nu} - \eta_{\mu\nu} \partial^2 H +
\partial_\mu \partial_\nu H
+ \eta_{\mu\nu} \partial^\alpha \partial^\beta H_{\alpha\beta}
- \partial_\sigma \partial_\mu H^\sigma{}_\nu
- \partial_\sigma \partial_\nu H^\sigma{}_\mu
\nonumber
\\&&\qquad\qquad
{}-m^2 H_{\mu\nu} + m^2 H \eta_{\mu\nu} = 0
\end{eqnarray}
contain $D$ primary constraints (expressions without second time
derivatives $\ddot H_{\mu\nu}$):
\bea
E_{00} &=& \Delta H_{ii} - \p_i\p_j H_{ij} - m^2 H_{ii} \equiv
\varphi_0^{(1)} \approx 0
\\
E_{0i} &=& \Delta H_{0i} +\p_i\dot H_{kk} -\p_k\dot H_{ki} - \p_i\p_k
H_{0k} - m^2 H_{0i} \equiv \varphi_i^{(1)} \approx 0 {.}
\eea
The remaining equations of motion $E_{ij}=0$ allow to define the
accelerations $\ddot H_{ij}$ in terms of $\dot H_{\mu\nu}$ and
$H_{\mu\nu}$. The accelerations $\ddot H_{00}$, $\ddot H_{0i}$ cannot
be expressed from the equations directly.

Conditions of conservation of the primary constraints in time
$\dot E_{0\mu}\approx 0$ lead to $D$ secondary constraints. On-shell
they are equivalent to
\beq
\varphi_\nu^{(2)} = \partial^\mu E_{\mu\nu} =
m^2 \partial_\nu H - m^2 \partial^\mu H_{\mu\nu} \approx 0
\label{con1}
\eeq

Conservation of $\varphi_i^{(2)}$ defines $D-1$ accelerations
$\ddot H_{0i}$ and conservation of $\varphi_0^{(2)}$ gives another
one constraint. It is convenient to choose it in the covariant form
by adding suitable terms proportional to the equations of motion:
\begin{eqnarray}
\varphi^{(3)}=
\partial^\mu \partial^\nu E_{\mu\nu}
+ \frac{m^2}{D-2} \eta^{\mu\nu} E_{\mu\nu}
=  H  m^4 \frac{D-1}{D-2} \approx 0
\label{con2}
\end{eqnarray}
Conservation of $\varphi^{(3)}$ gives one more constraint on initial
values
\beq
\varphi^{(4)} = - \dot H_{00} + \dot H_{kk} = \dot H \approx 0
\eeq
and from the conservation of this last constraint the acceleration
$\ddot H_{00}$ is defined.

Altogether there are $2D+2$ constraints on the initial values of
$\dot H_{\mu\nu}$ and $H_{\mu\nu}$. The theory contains the same local
dynamical degrees of freedom as the system (\ref{irred}) and
describes traceless and transverse symmetric tensor field of the
second rank.

Now if we want to construct a theory of massive spin 2 field on a
curved manifold first of all we should provide the same number of
propagating degrees of freedom as in the flat case. It means that new
equations of motion $E_{\mu\nu}$ should lead to exactly $2D+2$
constraints  and in the flat spacetime limit these constraints should
reduce to their flat counterparts. In addition to consistency with the
flat space limit any field theory should possess one more crucial
property connecting with causal propagation
\cite{zwanziger,zwanziger2}.  In general case interaction with
external fields changes light cones describing propagation of spin 2
massive field \cite{aragone} so causality may give another
restrictions on the theory.

Generalizing (\ref{actfield}) to curved spacetime we should substitute
all derivatives for the covariant ones and also we can add
non-minimal terms containing curvature tensor with some
dimensionless coefficients in front of them.  As a result, the most
general action for massive spin 2 field in curved spacetime quadratic
in derivatives and consistent with the flat limit should have the
form \cite{aragone}:
\begin{eqnarray}&&
S=\int d^D x\sqrt{-G} \biggl\{ \frac{1}{4} \nabla_\mu H \nabla^\mu H
-\frac{1}{4} \nabla_\mu H_{\nu\rho} \nabla^\mu H^{\nu\rho}
-\frac{1}{2} \nabla^\mu H_{\mu\nu} \nabla^\nu H
+\frac{1}{2} \nabla_\mu H_{\nu\rho} \nabla^\rho H^{\nu\mu}
\nonumber
\\&&
\qquad
{}+\frac{a_1}{2} R H_{\alpha\beta} H^{\alpha\beta}
+\frac{a_2}{2} R H^2
+\frac{a_3}{2} R^{\mu\alpha\nu\beta} H_{\mu\nu} H_{\alpha\beta}
+\frac{a_4}{2} R^{\alpha\beta} H_{\alpha\sigma} H_\beta{}^\sigma
+\frac{a_5}{2} R^{\alpha\beta} H_{\alpha\beta} H
\nonumber
\\&&
\qquad
{}- \frac{m^2}{4} H_{\mu\nu} H^{\mu\nu} + \frac{m^2}{4} H^2
 \biggr\}
\label{genact}
\end{eqnarray}
where $a_1, \ldots a_5$ are so far arbitrary dimensionless
coefficients, $R^\mu{}_{\nu\lambda\kappa}=\partial_\lambda
\Gamma^\mu_{\nu\kappa} -\ldots$,
$R_{\mu\nu}=R^\lambda{}_{\mu\lambda\nu}$.

Equations of motion
\begin{eqnarray}
E_{\mu\nu}&=&\nabla^2 H_{\mu\nu} - G_{\mu\nu} \nabla^2 H +
\nabla_\mu \nabla_\nu H
+ G_{\mu\nu} \nabla^\alpha \nabla^\beta H_{\alpha\beta}
- \nabla_\sigma \nabla_\mu H^\sigma{}_\nu
- \nabla_\sigma \nabla_\nu H^\sigma{}_\mu
\nonumber
\\&&
{}+2a_1 R H_{\mu\nu}
+2a_2 G_{\mu\nu} R H
+2a_3 R_\mu{}^\alpha{}_\nu{}^\beta H_{\alpha\beta}
+a_4 R_\mu{}^\alpha H_{\alpha\nu}
+a_4 R_\nu{}^\alpha H_{\alpha\mu}
\nonumber
\\&&
{}+a_5 R_{\mu\nu} H
+a_5 G_{\mu\nu} R^{\alpha\beta} H_{\alpha\beta}
-m^2 H_{\mu\nu} + m^2 H G_{\mu\nu} \approx 0
\label{lagreq}
\end{eqnarray}
contain second time derivatives of $H_{\mu\nu}$ in the following
way:
\bea
E_{00} &=& (G^{mn}-G_{00}G^{00}G^{mn}+G_{00}G^{0m}G^{0n}) \nabla_0
\nabla_0 H_{mn} + O(\nabla_0) {,}
\nonumber\\
E_{0i} &=&
(-G_{0i}G^{00}G^{mn} + G_{0i}G^{0m}G^{0n} - G^{0m}\delta^n_i)
\nabla_0 \nabla_0 H_{mn} + O(\nabla_0) {,}
\nonumber\\
E_{ij} &=&
(G^{00}\delta^m_i\delta^n_j - G_{ij}G^{00}G^{mn} +
G_{ij}G^{0m}G^{0n}) \nabla_0 \nabla_0 H_{mn} + O(\nabla_0) {.}
\eea
So we see that accelerations $\ddot H_{00}$ and $\ddot H_{0i}$ again
(as in the flat case) do not enter the equations of motion while
accelerations $\ddot H_{ij}$ can be expressed through $\dot
H_{\mu\nu}$, $H_{\mu\nu}$ and their spatial derivatives.

There are $D$ linear combinations of the equations of motion which
do not contain second time derivatives and so represent primary
constraints of the theory:
\be
\varphi^{(1)}_\mu = E^0{}_\mu = G^{00} E_{0\mu} + G^{0j} E_{j\mu}
\label{phi1}
\ee
Now one should calculate time derivatives of these constraints and
define secondary ones. In order to do this in a covariant form
we can add to the time derivative of $\varphi^{(1)}_\mu$ any linear
combination of equations of motion and primary constraints. So we
choose the secondary constraints in the following way:
\bea
\varphi^{(2)}_\mu &=& \nabla^\alpha E_{\alpha\mu} =
\dot\varphi^{(1)}_\mu + \p_i E^i{}_\mu
+ \Gamma^\alpha_{\alpha0}\varphi^{(1)}_\mu
+ \Gamma^\alpha_{\alpha i} E^i{}_\mu
- \Gamma^\sigma_{\mu 0}\varphi^{(1)}_\sigma
- \Gamma^\sigma_{\mu i} E^i{}_\sigma
\nonumber\\
&=&
(2a_1 R-m^2) \nabla^\mu H_{\mu\nu}
+(2a_2 R+m^2) \nabla_\nu H
+2a_3 R^{\mu\alpha}{}_\nu{}^\beta \nabla_\mu H_{\alpha\beta}
+a_4 R^{\mu\alpha} \nabla_\mu H_{\alpha\nu}
\nonumber\\&&{}
+(a_4-2) R^\alpha{}_\nu \nabla^\mu H_{\alpha\mu}
+a_5 R^{\alpha\mu} \nabla_\nu H_{\alpha\mu}
+(a_5+1) R^\alpha{}_\nu \nabla_\alpha H
\nonumber\\&&{}
+ (2a_1+\frac{a_4}{2}) H_{\alpha\nu} \nabla^\alpha R
+ (2a_2+\frac{a_5}{2}) H \nabla_\nu R
\nonumber\\&&{}
+ H_{\alpha\beta} \biggr[ (2a_3+a_5+1) \nabla_\nu R^{\alpha\beta}
+ (a_4-2a_3-2) \nabla^\alpha R^\beta{}_\nu \biggl]
\label{phi2}
\eea
At the next step conservation of these $D$ secondary constraints
should lead to one new constraint and to expressions for $D-1$
accelerations $\ddot H_{0i}$. This means that the constraints
(\ref{phi2}) should contain the first time derivatives $\dot
H_{0\mu}$ through the matrix with the rank $D-1$:
\bea
\varphi^{(2)}_0 &=&
A \; \dot H_{00} + B^j \dot H_{0j} + \ldots
\nonumber\\
\varphi^{(2)}_i &=&
C_i \dot H_{00} + D_i{}^j \dot H_{0j} + \ldots
\label{velocities}
\eea
\be
\mbox{rank}\; \hat\Phi_\mu{}^\nu \equiv
\mbox{rank} \left|\left|
\begin{array}{cc}
A& B^j \\ C_i & D_i{}^j
\end{array}
\right|\right| = D-1
\label{rank}
\ee
In the flat spacetime we had the matrix
\be
\hat\Phi_\mu{}^\nu =
\left|\left|
\begin{array}{cc}
0& 0 \\ 0 & m^2 \delta_i^j
\end{array}
\right|\right|
\ee
In the curved case the explicit form of this matrix elements in the
constraints (\ref{phi2}) is:
\bea
A &=& R G^{00} (2a_1+2a_2) + R^{00} (a_4+a_5)
           + R^0{}_0 G^{00} (a_4+a_5-1)
\nonumber\\
B^j &=& m^2 G^{0j} + R G^{0j} (2a_1+4a_2) + 2a_3 R^{0j}{}_0{}^0
      + R^j{}_0 G^{00} (a_4-2)
\nonumber\\&&{}
      + R^{0j} (a_4+2a_5) + R^0{}_0 G^{0j} (a_4+2a_5)
\nonumber\\
C_i &=& R^0{}_i G^{00} (a_4+a_5-1)
\nonumber\\
D_i{}^j &=&{} - m^2 G^{00} \delta_i^j + 2a_1 RG^{00} \delta_i^j
 + 2a_3 R^{0j}{}_i{}^0 + a_4 R^{00} \delta_i^j
\nonumber\\&&{}
 + (a_4-2) R^j{}_i G^{00} + (a_4+2a_5) R^0_i G^{0j}
\label{elements}
\eea

At this stage the restrictions that consistency imposes on the
non-minimal couplings and on the external gravitational field reduce
to the requirements that the above matrix elements give
$\det\hat\Phi=0$ while $\det D_i{}^j \neq 0$.

One way to fulfill these requirements is to impose the
following restriction on the external gravitational fields:
\be
R_{\mu\nu}=\frac{1}{D}G_{\mu\nu}R \; {.}
\label{einst}
\ee
It means that one considers only Einstein spacetimes \cite{petrov}
representing solutions of vacuum Einstein equations with cosmological
constant. In these spacetimes the scalar curvature $R$ is constant as
follows from the Bianchi identity
$\nabla^\mu R_{\mu\nu} = \frac{1}{2} \nabla_\nu R$
but the Weyl tensor part of the curvature tensor can be arbitrary.

If the Einstein equation (\ref{einst}) for external gravity is
fulfilled the coefficients $a_4$, $a_5$ in the lagrangian
(\ref{genact}) are absent and the matrix $\hat\Phi$ takes the form:
\be
\hat\Phi_\mu{}^\nu=
\left|\left|
\begin{array}{c|c}
R G^{00}(2a_1+2a_2-\frac{\d 1}{\d D})
& R G^{0j}(2a_1+4a_2) + 2a_3R^{0j}{}_0{}^0 + m^2 G^{0j} \\
~&~\\
\hline
~&~\\
0& 2a_3 R^{0j}{}_i{}^0 +
 R G^{00} \delta_i^j (2a_1-\frac{\d 2}{\d D}) - m^2 G^{00} \delta_i^j
\end{array}
\right|\right|
\ee
The simplest way to make the rank of this matrix to be equal to $D-1$
is provided by the following choice of the coefficients:
\be
2a_1 + 2a_2 -\frac{1}{D} = 0, \qquad a_3 = 0, \qquad
2R \biggl(a_1-\frac{1}{D}\biggr) - m^2 \neq 0 {.}
\ee

As a result, we have one-parameter family of theories:
\bea
&&
a_1 =\frac{\xi}{D}, \quad a_2 = \frac{1-2\xi}{2D},\quad
a_3=0,\quad a_4=0, \quad a_5 = 0
\nonumber\\&&
R_{\mu\nu}=\frac{1}{D}G_{\mu\nu}R , \qquad
\frac{2(1-\xi)}{D} R + m^2 \neq 0 {.}
\eea
with $\xi$ an arbitrary real number.

The action in this case takes the form
\bea
&&
S=\int d^D x\sqrt{-G} \biggl\{ \frac{1}{4} \nabla_\mu H \nabla^\mu H
-\frac{1}{4} \nabla_\mu H_{\nu\rho} \nabla^\mu H^{\nu\rho}
-\frac{1}{2} \nabla^\mu H_{\mu\nu} \nabla^\nu H
+\frac{1}{2} \nabla_\mu H_{\nu\rho} \nabla^\rho H^{\nu\mu}
\nonumber
\\&&
\qquad\qquad
+\frac{\xi}{2D} R H_{\mu\nu} H^{\mu\nu} +\frac{1-2\xi}{4D} R H^2
- \frac{m^2}{4} H_{\mu\nu} H^{\mu\nu} + \frac{m^2}{4} H^2
 \biggr\} {.}
\label{curvact}
\eea
and the corresponding equations of motion are
\bea
&&
E_{\mu\nu}=\nabla^2 H_{\mu\nu} - G_{\mu\nu} \nabla^2 H +
\nabla_\mu \nabla_\nu H
+ G_{\mu\nu} \nabla^\alpha \nabla^\beta H_{\alpha\beta}
- \nabla_\sigma \nabla_\mu H^\sigma{}_\nu
- \nabla_\sigma \nabla_\nu H^\sigma{}_\mu
\nonumber
\\&&\qquad\qquad
+ \frac{2\xi}{D} R H_{\mu\nu} + \frac{1-2\xi}{D} RH G_{\mu\nu}
-m^2 H_{\mu\nu} + m^2 H G_{\mu\nu} = 0
\label{cons_eq}
\eea
The secondary constraints built out of them are
\be
\varphi^{(2)}_\mu=\nabla^\alpha E_{\alpha\mu} =
(\nabla_\mu H -\nabla^\alpha H_{\mu\alpha})
\biggl( m^2+\frac{2(1-\xi)}{D}R\biggr)
\ee
and the matrix $\hat\Phi$ looks like
\be
\hat\Phi_\mu{}^\nu =
\biggl( m^2+\frac{\d 2(1-\xi)}{\d D}R\biggr)
\left|\left|
\begin{array}{c|c}
0& G^{0j}  \\
\hline
0 & - G^{00}\delta_i^j
\end{array}
\right|\right|
\ee
Just like in the flat case, in this theory the conditions
$\dot \varphi^{(2)}_i\approx 0$ define the accelerations $\ddot H_{0i}$
and the condition $\dot\varphi^{(2)}_0\approx 0$ after excluding
$\ddot H_{0i}$ gives a new constraint, i.e. the acceleration $\ddot
H_{00}$ is not defined at this stage.

To define the new constraint in a covariant form we use the
following linear combination of $\dot\varphi^{(2)}_\mu$, equations of
motion, primary and secondary constraints:
\bea
\varphi^{(3)} &=&
\frac{m^2}{D-2} G^{\mu\nu} E_{\mu\nu}
+ \nabla^\mu \nabla^\nu E_{\mu\nu}
+ \frac{2(1-\xi)}{D(D-2)} R G^{\mu\nu} E_{\mu\nu} =
\nonumber
\\&=& H \frac{1}{D-2} \biggl( \frac{2(1-\xi)}{D} R + m^2 \biggr)
\biggl( \frac{D+2\xi(1-D)}{D} R + m^2 (D-1)\biggr) \approx 0 {.}
\eea
Requirement of its conservation leads to one more constraint
\bea
\dot \varphi^{(3)} \sim \dot H  \quad\Longrightarrow\quad
\varphi^{(4)} = \dot H \approx 0 {.}
\eea
The last acceleration $\ddot H_{00}$ is expressed from the
condition $\dot\varphi^{(4)}\approx 0$.

Using the constraints for simplifying the equations of motion we see
that the original equations are equivalent to the following system:
\bea
&&\nabla^2 H_{\mu\nu}
+ 2 R^\alpha{}_\mu{}^\beta{}_\nu H_{\alpha\beta}
+ \frac{2(\xi-1)}{D} R H_{\mu\nu} - m^2 H_{\mu\nu} = 0 {,}
\nonumber\\&&
H^\mu{}_\mu=0 {,} \qquad\qquad\dot H^\mu{}_\mu=0 {,} \qquad\qquad
\nabla^\mu H_{\mu\nu} = 0 {,}
\label{curv_shell}
\\&&
G^{00}\n_0\n_i H^i{}_\nu - G^{0i}\n_0\n_i H^0{}_\nu
- G^{0i}\n_i\n_0 H^0{}_\nu - G^{ij}\n_i\n_j H^0_\nu
-2R^{\alpha0\beta}{}_\nu H_{\alpha\beta}
\nonumber
\\&&{}\qquad\qquad\qquad\qquad
- \frac{2(\xi-1)}{D} RH^0{}_\nu + m^2 H^0{}_\nu = 0 {.}
\nonumber
\eea
The last expression represents $D$ primary constraints.

For any values of $\xi$ the theory describes the same number of
degrees of freedom as in the flat case - the symmetric, covariantly
transverse and traceless tensor. $D$ primary constraints guarantees
conservation of the transversality conditions in time.

Now we turn to the discussion of causality for the spin 2 field
equations. Analogous problem with dynamical gravity was investigated
in \cite{aragone} (see also \cite{zwanziger,zwanziger2} for
causality problem in electromagnetic background).  In general, when
one has a system of differential equations for a set of fields
$\phi^B$ (to be specific, let us say about second order equations)
\be
M_{AB}{}^{\mu\nu}\p_\mu \p_\nu \phi^B + \ldots = 0 {,} \qquad
\mu,\nu=0,\ldots,D-1
\ee
the following definitions are used. A characteristic matrix is the
matrix function of $D$ arguments $n_\mu$ built out of the
coefficients at the second derivatives in the equations:
$
M_{AB}(n) = M_{AB}{}^{\mu\nu} n_\mu n_\nu {.}
$
A characteristic equation is
$
\det M_{AB} (n) = 0 {.}
$
A characteristic surface is the surface $S(x)=const$ where
$\p_\mu S(x)=n_\mu$.

If for any $n_i$ ($i=1,\ldots,D-1$) all solutions of the
characteristic equation $n_0(n_i)$ are real then the system of
differential equations is called hyperbolic and describes
propagation of some wave processes. The hyperbolic system is called
causal if there is no timelike vectors among solutions $n_\mu$ of the
characteristic equations. Such a system describes propagation with a
velocity not exceeding the speed of light. If there exist timelike
solutions for $n_\mu$ then the corresponding characteristic surfaces
are spacelike which violates causality.

In the flat spacetime the equations for the spin 2 field (\ref{irred})
lead to the characteristic equation
\be
\det M(n) = (n^2)^{D(D+1)/2}
\ee
which has 2 multiply degenerate roots:
\be
-n_0^2+n_i^2 = 0, \qquad n_0 = \pm \sqrt{n_i^2} {.}
\ee
The solutions for $n_\mu$ are real and null hence the equations are
hyperbolic and causal.

Now consider the curved spacetime generalization. If we tried to use
the equations of motion in the original lagrangian form
(\ref{cons_eq}) then the characteristic matrix
\be
M_{\mu\nu}{}^{\lambda\kappa} (n) =
\delta_{(\mu\nu)}{}^{(\lambda\kappa)} n^2
- G_{\mu\nu} G^{\lambda\kappa}n^2
+ G^{\lambda\kappa} n_\mu n_\nu
+ G_{\mu\nu} n^\lambda n^\kappa
- \delta_\nu^{(\kappa} n^{\lambda)} n_\mu
- \delta_\mu^{(\kappa} n^{\lambda)} n_\nu
\ee
would be degenerate. This fact can be seen from the relation
\be
n^\mu M_{\mu\nu}{}^{\lambda\kappa} (n) \equiv 0
\ee
which means that any symmetric tensor of the form $n_{(\mu}t_{\nu)}$
(with $t_\nu$ an arbitrary vector) represents a ``null vector'' for
the matrix $M(n)$  and therefore $\det M=0$.

After having used the constraints we obtain the equations of motion
written in the form (\ref{curv_shell}) and the characteristic
matrix becomes non-degenerate:
\be
M_{\mu\nu}{}^{\lambda\kappa} (n) =
\delta_{\mu\nu}{}^{\lambda\kappa} n^2, \qquad
n^2 = G^{\alpha\beta} n_\alpha n_\beta {.}
\ee
The characteristic cones remains the same as in the flat case. At any
point $x_0$ we can choose locally
$G^{\alpha\beta}(x_0)=\eta^{\alpha\beta}$
and then
\be
\left. n^2 \right|_{x_0} = - n_0^2 + n_i^2
\ee
Just like in the flat case the equations are hyperbolic and causal.

So we demonstrated that in Einstein spacetimes spin 2 massive field
can be consistently described by a one-parameter family of theories
(\ref{curvact}). For any value of the parameter the corresponding
equations describe the correct number of degrees of freedom which
propagate causally. Our lagrangian for the spin 2 field in curved
spacetime is the most general known so far, in all previous works
only the theories with specific values of the parameter $\xi$ were
considered \cite{bengt,ovrut}.

The next natural step would consist in building a theory describing
dynamics of both gravity and massive spin 2 field. In such a theory
in addition to dynamical equations for the massive spin 2 field one
would have dynamical equations for gravity with the energy-momentum
tensor constructed out of spin 2 field components.  The analysis of
consistency then changes and one needs to have correct number of
constraints and causality for both fields interacting with each
other \cite{aragone}.

The only known consistent system of a higher spin field interacting
with dynamical gravity is the theory of massless helicity 3/2 field,
i.e. supergravity \cite{sugra} (see also the book \cite{book}). In
that case consistency with dynamical gravity requires four-fermion
interaction.  If a consistent description of spin 2 field interacting
with dynamical gravity exists it may also require some non-trivial
modification of the lagrangian.  At least, it is known that
lagrangians quadratic in spin 2 field do not provide such a
consistency \cite{aragone}. In the Section~4 we will describe a
possible way of consistent description of the spin 2 field on
arbitrary gravitational background which is given by representation
of the lagrangian in the form of infinite series in inverse mass.

\section{Coupling to background scalar field}

Now we investigate a possibility to generalize the above
analysis for the case of spin 2 massive field interacting not only
with background gravity but also with a scalar dilaton field. This
set of fields arises naturally in string theory which contains
dilaton field $\phi(x)$ as one of its massless excitations.

Writing a general action similar to (\ref{curvact}) for this system
one should take into account all possible new terms
with derivatives of $\phi(x)$ and also containing arbitrary
factors $f(\phi)$ without derivatives of the dilaton fields. For
example, string effective action can contain in various terms the
factors $e^{k\phi}$, $k=const$. We will consider here the class of
actions for the field $H_{\mu\nu}$ where all these factors can be
absorbed to the metric $G_{\mu\nu}$ by a conformal rescaling.

The most general action of this type is
\begin{equation}
S=S_G+S_\phi {,}
\label{actdil}
\end{equation}
where $S_G$ is the general action without dependence on scalar field
(\ref{genact}) and $S_\phi$ can contain (up to total derivatives) ten
new terms:
\begin{eqnarray}
S_\phi &=& \int d^D x\sqrt{-G}
\biggl\{
\frac{c_1}{2} H^{\mu\nu}\nabla_\alpha H_{\mu\nu}\nabla^\alpha\phi
+\frac{c_2}{2} H \nabla_\alpha H \nabla^\alpha\phi
+\frac{c_3}{2} H^{\mu\nu} \nabla_\mu H_{\alpha\nu} \nabla^\alpha \phi
\nonumber\\&& {}
+\frac{c_4}{2} H \nabla^\mu H_{\alpha\mu} \nabla^\alpha \phi
+\frac{c_5}{2} H_{\mu\nu} \nabla^\nu H  \nabla^\mu \phi
+\frac{c_6}{2}  H_{\alpha\beta} \nabla^\mu H_{\mu}{}^\beta\nabla^\alpha \phi
+\frac{c_7}{2} H_{\mu\alpha} H_\nu{}^\alpha \nabla^\mu \phi \nabla^\nu \phi
\nonumber\\&&{}
+\frac{c_8}{2} H H_{\mu\nu} \nabla^\mu \phi \nabla^\nu \phi
+\frac{c_9}{2}  H_{\mu\nu} H^{\mu\nu} (\nabla \phi)^2
+\frac{c_{10}}{2} H^2 (\nabla \phi)^2
\biggr\}
\end{eqnarray}
We consider $\phi$ as a background field on the same footing with the
metric $G_{\mu\nu}$.

What values of coupling parameters $c_1,\ldots,c_{10}$ are permissible
and how does the condition on the background (\ref{einst}) change in
the presence of $\phi$? To answer these questions one should repeat
the analysis of the previous section for the action (\ref{actdil})
calculating all constraints of the theory.

The equations of motion are:
\begin{eqnarray}
E_{\mu\nu}&=&\nabla^2 H_{\mu\nu} - G_{\mu\nu} \nabla^2 H +
\nabla_\mu \nabla_\nu H
+ G_{\mu\nu} \nabla^\alpha \nabla^\beta H_{\alpha\beta}
- \nabla_\sigma \nabla_\mu H^\sigma{}_\nu
- \nabla_\sigma \nabla_\nu H^\sigma{}_\mu
\nonumber\\&&
{}+2a_1 R H_{\mu\nu}
+2a_2 G_{\mu\nu} R H
+2a_3 R_\mu{}^\alpha{}_\nu{}^\beta H_{\alpha\beta}
+a_4 R_\mu{}^\alpha H_{\alpha\nu}
+a_4 R_\nu{}^\alpha H_{\alpha\mu}
\nonumber\\&&
{}+a_5 R_{\mu\nu} H
+a_5 G_{\mu\nu} R^{\alpha\beta} H_{\alpha\beta}
+\frac{c_3-c_6}{2} (\nabla_\mu H_{\alpha\nu} \nabla^\alpha \phi +
                      \nabla_\nu H_{\alpha\mu} \nabla^\alpha \phi)
\nonumber\\&&
+\frac{c_6-c_3}{2} (\nabla^\alpha H_{\alpha\nu} \nabla_\mu \phi +
                      \nabla^\alpha H_{\alpha\mu} \nabla_\nu \phi)
+\frac{c_5-c_4}{2} (\nabla_\nu H \nabla_\mu \phi +
                      \nabla_\mu H \nabla_\nu \phi)
\nonumber\\&&
+(c_4-c_5) G_{\mu\nu} \nabla^\beta H_{\alpha\beta} \nabla^\alpha \phi
-\frac{c_3+c_6}{2} ( H_{\alpha\nu} \nabla^\alpha \nabla_\mu \phi
                   + H_{\alpha\mu} \nabla^\alpha \nabla_\nu \phi)
\nonumber\\&&
-c_1  H_{\mu\nu} \nabla^2 \phi - c_2 G_{\mu\nu} H \nabla^2 \phi
-c_4  H \nabla_\mu \nabla_\nu \phi
-c_5 G_{\mu\nu} H_{\alpha\beta} \nabla^\alpha \nabla^\beta \phi
\nonumber\\&&
+c_7 ( H_{\nu\alpha} \nabla_\mu \phi \nabla^\alpha \phi
       + H_{\mu\alpha} \nabla_\nu \phi \nabla^\alpha \phi )
+c_8  H \nabla_\mu \phi \nabla_\nu \phi
+c_8  G_{\mu\nu} H_{\alpha\beta} \nabla^\alpha \phi \nabla^\beta \phi
\nonumber\\&&
+2 c_9  H_{\mu\nu} (\nabla \phi)^2
+2 c_{10} G_{\mu\nu} H (\nabla \phi)^2
-m^2 H_{\mu\nu} + m^2 H G_{\mu\nu} = 0
\end{eqnarray}

Introducing of background dilaton does not change the second
derivatives terms in the equations and so
$\varphi^{(1)}_\mu=E^0{}_\mu\approx 0$ are again primary constraints.
Conditions of their conservation should give $D$ secondary
constraints:
\bea
\varphi^{(2)}_0 &=& \n^\mu E_{\mu 0}  =
\frac{1}{2} (c_3-c_6+c_4-c_5) G^{00}\n_k\phi (G^{0k}\ddot
H_{00}+2G^{kj}\ddot H_{0j}) + \ldots {,}
\nonumber\\
\varphi^{(2)}_i &=& \n^\mu E_{\mu i}  =
- \frac{1}{2} (c_3-c_6+c_4-c_5) G^{00}\n_i\phi
(G^{00}\ddot H_{00}+2G^{0j}\ddot H_{0j}) + \ldots
\eea
Hence we should impose the restriction
\be
c_3-c_6+c_4-c_5 = 0 {.}
\ee
in order to cancel second time derivatives in $\varphi^{(2)}_\mu$. Note
that without dilaton couplings these expressions do not contain
second derivatives at all and represent constraints for any values of
non-minimal couplings with gravity (\ref{phi1}).

We will restrict ourselves to even simpler particular class of the
theories with
\be
c_3=c_6, \qquad c_4=c_5 {.}
\label{dilprim}
\ee
Then the secondary constraints contain first time derivatives of
$H_{0\mu}$ with the following coefficients (\ref{velocities}):
\bea
A &=& G^{00} \Bigl[ 2(a_1+a_2)R -(c_1+c_2)\n^2\phi
+2(c_9+c_{10}) (\n\phi)^2 \Bigr]
\nonumber\\&&{}
+ (a_4+a_5)R^{00} -(c_3+c_4)\n^0\n^0\phi + (c_7+c_8)\n^0\phi\n^0\phi
\nonumber\\&&{}
+  G^{00}\Bigl[ (a_4+a_5-1) R^0{}_0 -(c_3+c_4)\n_0\n^0\phi +
(c_7+c_8)\n_0\phi\n^0\phi \Bigr]
\nonumber\\
B^j &=& 2a_3 R^{0j}{}_0{}^0 +
G^{0j} \Bigl[ m^2  + (2a_1+4a_2) R  -(c_1+2c_2)\n^2\phi +
(2c_9+4c_{10}) (\n\phi)^2 \Bigr]
\nonumber\\&&{}
+  (a_4+2a_5) R^{0j} - (c_3+2c_4) \n^0\n^j\phi +
   (c_7+2c_8)\n^0\phi\n^j\phi
\nonumber\\&&{}
   + G^{0j}\Bigl[ (a_4+2a_5)R^0{}_0 - (c_3+2c_4) \n_0\n^0\phi +
   (c_7+2c_8)\n_0\phi\n^0\phi  \Bigr]
\nonumber\\&&{}
    +G^{00} \Bigl[ (a_4-2)R^j{}_0 -c_3\n_0\n^i\phi
   + c_7\n_0\phi \n^i\phi \Bigr]
\nonumber\\
C_i &=&  G^{00}\Bigl[ (a_4+a_5-1) R^0{}_i -(c_3+c_4)\n_i\n^0\phi +
(c_7+c_8)\n_i\phi\n^0\phi \Bigr]
\nonumber\\
D_i{}^j &=& 2a_3 R^{0j}{}_i{}^0 +
G^{00} \delta_i^j \Bigl[ -m^2 + 2a_1 R -c_1\n^2\phi+ 2c_9(\n\phi)^2\Bigr]
\nonumber\\&&{}
   + \delta_i^j \Bigl[ a_4 R^{00} -c_3\n^0\n^0\phi
      +\n^0\phi \n^0\phi \Bigr]
\nonumber\\&&{}
 + G^{00} \Bigl[ (a_4-2) R^j{}_i -c_3\n_i\n^j\phi
       +c_7\n_i\phi \n^j\phi \Bigr]
\nonumber\\&&{}
 + G^{0j} \Bigl[ (a_4+2a_5) R^0_i -(c_3+2c_4) \n_i\n^0\phi
 + (c_7+2c_8) \n_i\phi \n^0\phi \Bigr]
\label{dil_matrix}
\eea

The simplest way to make the rank of the matrix $\hat\Phi$ (\ref{rank})
with the elements (\ref{dil_matrix}) to be equal to $D-1$ consists in
the choice
\bea
&& R_{\mu\nu}=\frac{1}{D} RG_{\mu\nu} \; , \qquad
2a_1+2a_2-\frac{1}{D}=0 \; , \qquad     a_3=a_4=a_5=0 \; ,
\nonumber\\&&
c_1=-c_2 \; , \qquad c_3=-c_4 \; , \qquad c_7=-c_8 \; ,
\qquad c_9=-c_{10} \; .
\eea
supplemented by (\ref{dilprim})

In this case consistent action for the spin 2 field
interacting with background gravitational and scalar fields
contains five arbitrary constant parameters:
\begin{eqnarray}
&& S=\int d^D x\sqrt{-G}
\biggl\{ \frac{1}{4} \nabla_\mu H \nabla^\mu H -\frac{1}{4}
\nabla_\mu H_{\nu\rho} \nabla^\mu H^{\nu\rho} -\frac{1}{2} \nabla^\mu
H_{\mu\nu} \nabla^\nu H +\frac{1}{2} \nabla_\mu H_{\nu\rho}
\nabla^\rho H^{\nu\mu}
\nonumber\\&& \qquad{}
+\frac{\xi}{2D} R H_{\mu\nu}
H^{\mu\nu} +\frac{1-2\xi}{4D} R H^2
+\frac{\zeta_1}{2}\nabla^\alpha\phi\nabla_\alpha H_{\mu\nu}H^{\mu\nu}
-\frac{\zeta_1}{2}\nabla^\alpha\phi\nabla_\alpha H H
\nonumber\\&& \qquad{}
+\frac{\zeta_2}{2}\nabla^\alpha \phi \nabla_\mu H_{\alpha\nu} H^{\mu\nu}
-\frac{\zeta_2}{2}\nabla^\alpha \phi \nabla^\mu H_{\alpha\mu} H
-\frac{\zeta_2}{2}\nabla^\mu \phi \nabla^\nu H H_{\mu\nu}
+\frac{\zeta_2}{2}\nabla^\alpha \phi \nabla^\mu H_{\mu}{}^\beta H_{\alpha\beta}
\nonumber\\&&
\qquad{}
+\frac{\zeta_3}{2}\nabla^\mu \phi \nabla^\nu \phi H_{\mu\alpha} H_\nu{}^\alpha
-\frac{\zeta_3}{2}\nabla^\mu \phi \nabla^\nu \phi H_{\mu\nu} H
+\frac{\zeta_4}{2} (\nabla \phi)^2 H_{\mu\nu} H^{\mu\nu}
-\frac{\zeta_4}{2} (\nabla \phi)^2 H^2
\nonumber\\&&
\qquad{}
- \frac{m^2}{4} H_{\mu\nu} H^{\mu\nu} + \frac{m^2}{4} H^2
 \biggr\}
\label{dilact}
\end{eqnarray}
and conservation conditions for the secondary constraints
$\varphi^{(2)}_\mu=\n^\nu E_{\mu\nu}$ can be used for building one
new constraint:
\bea
\varphi^{(3)}&=&\n^\mu\n^\nu E_{\mu\nu} +
(\zeta_3\n^\mu\phi\n^\nu\phi - \zeta_2 \n^\mu\n^\nu\phi ) E_{\mu\nu}
\nonumber\\&&{}\qquad
+\biggl( m^2 + \frac{2(1-\xi)}{D}R -(\zeta_3+2\zeta_4)(\n\phi)^2 +
(\zeta_1+\zeta_2) \n^2\phi \biggr) \frac{1}{D-2} G^{\mu\nu} E_{\mu\nu}
\nonumber\\
&=& \n_\alpha H_{\mu\nu}
           \Bigl[
    - 2\zeta_2 R^{\mu\alpha\nu\beta} \n_\beta\phi
    + 2\zeta_3 \n^\mu \n^\nu\phi \n^\alpha\phi
    - 2\zeta_3 \n^\mu \n^\alpha\phi \n^\nu\phi
           \Bigr]
\nonumber\\&&{}
+ (\n_\nu H - \n^\mu H_{\mu\nu})
           \Bigl[
   2(\zeta_1+\zeta_2) \n^\nu\n^2\phi
   + \frac{2\xi_2}{D} R \n^\nu\phi
   -2\zeta_3 \n^\nu\phi \n^2\phi
\nonumber\\&&{}
   -2(\zeta_3+4\zeta_4) \n^\nu\n^\alpha\phi \n_\alpha\phi
           \Bigr]
+ H_{\mu\nu} f^{\mu\nu} (\phi) + H f(\phi) {.}
\label{dilcaus}
\eea
Here $f^{\mu\nu}$ and $f$ are functions of the scalar field $\phi$
and its derivatives, we will not write the explicit form of them here.
The important fact about this constraint is that it does not contain
the time derivative of the field component $H_{00}$. Hence, the
conservation condition $\dot\varphi^{(3)}\approx 0$ does not contain
the acceleration $\ddot H_{00}$ and leads to another one constraint
$\varphi^{(4)}$ providing thus the correct total number of
constraints in the theory.  The acceleration $\ddot H_{00}$ is
defined only from the condition $\dot\varphi^{(4)}\approx 0$. For any
values of the coupling parameters  the theory (\ref{dilact})
describes correct number of degrees of freedom.

Causality in presence of dilaton may be violated even in
the flat spacetime. It follows from the fact that in general the
constraints $\varphi^{(2)}_\mu$ and $\varphi^{(3)}$ cannot be solved
algebraically with respect to the trace $H$ and the longitudinal part
$\n^\mu H_{\mu\nu}$ and used for cancelling the corresponding terms
in the equations of motion. As a result, the characteristic matrix
differs signifigantly from its flat counterpart and the characteristic
equation may possess new non-trivial solutions spoiling causality
for some values of the parameters $\zeta$, $\xi$.  We postpone a
detailed study of the causality in the presence of dilaton for a
separate publication.

In the rest of the paper we will be considering only pure
gravitational background when the scalar field $\phi$ is absent.

\section{Consistent equations in arbitrary background}

In previous sections we analyzed a possibility of consistent
description of the spin 2 field on arbitrary Einstein manifold. Now
we will describe another possibility which allows to remove any
restrictions on the external gravitational background by means of
considering a lagrangian in the form of infinite series in inverse
mass $m$. Existence  of dimensionful mass parameter $m$ in the theory
let us construct a lagrangian with terms of arbitrary orders
in curvature multiplied by the corresponding powers of $1/m^2$:
\begin{eqnarray}&&
S_H =\int d^D x\sqrt{-G} \biggl\{ \frac{1}{4} \nabla_\mu H \nabla^\mu H
-\frac{1}{4} \nabla_\mu H_{\nu\rho} \nabla^\mu H^{\nu\rho}
-\frac{1}{2} \nabla^\mu H_{\mu\nu} \nabla^\nu H
+\frac{1}{2} \nabla_\mu H_{\nu\rho} \nabla^\rho H^{\nu\mu}
\nonumber
\\&&
\qquad
{}+\frac{a_1}{2} R H_{\alpha\beta} H^{\alpha\beta}
+\frac{a_2}{2} R H^2
+\frac{a_3}{2} R^{\mu\alpha\nu\beta} H_{\mu\nu} H_{\alpha\beta}
+\frac{a_4}{2} R^{\alpha\beta} H_{\alpha\sigma} H_\beta{}^\sigma
+\frac{a_5}{2} R^{\alpha\beta} H_{\alpha\beta} H
\nonumber\\&&\qquad{}
+\frac{1}{m^2} ( R \nabla H \nabla H + R H \nabla\nabla H + R R H H)
+\frac{1}{m^4} (R R \nabla H \nabla H + R R H \nabla\nabla H
\nonumber\\&&\qquad{}
+ R \nabla R H \nabla H + R \nabla\nabla R H H + R R R H H)
+ O\Bigl(\frac{1}{m^6}\Bigr)
\nonumber\\&&\qquad{}
- \frac{m^2}{4} H_{\mu\nu} H^{\mu\nu} + \frac{m^2}{4} H^2
 \biggr\}
\label{higher}
\end{eqnarray}
Actions of this kind are expected to arise naturally in string
theory where the role of mass parameter is played by string
tension $m^2=1/\alpha'$ and perturbation theory in $\alpha'$ will
give for background fields effective actions of the form
(\ref{higher}) \footnote{In fact, string theory should lead to even
more general effective actions than (\ref{higher}) since in the
higher $\alpha'$ corrections higher derivatives of all background
fields should appear.}.

Possibility of constructing consistent equations for massive higher
spin fields as series in curvature was recently studied in
\cite{klish2} where such equations were derived
in particular case of symmetrical Einstein spaces in linear in
curvature order.

Here we will demonstrate that requirement of consistency with the
flat spacetime limit can be fulfilled perturbatively in $1/m^2$
for arbitrary gravitational background at least in the lowest order.
We will use the same general scheme of calculating lagrangian
constraints as in the previous sections. The only difference is that
each condition will be considered perturbatively and can be solved
separately in each order in $1/m^2$.

Primary constraints in the theory described by the action
(\ref{higher}) should be given by the equations $E^0{}_\mu\approx 0$.
Requirement of absence of second time derivatives in these equations
will give some restrictions on coefficients in higher orders in
$1/m^2$, for example, in terms like $R\n H\n H$.

Secondary constraints in the lowest order in $1/m^2$ were already
calculated in the Section~2 (\ref{phi2}). Consistency with the flat
spacetime limit requires existence of one additional constraint among
conservation conditions of these secondary constraints. So we should
calculate the rank of the matrix
\be
\hat\Phi_\mu{}^\nu =
\left|\left|
\begin{array}{cc}
A& B^j \\ C_i & D_i{}^j
\end{array}
\right|\right|
\ee
with the elements in the lowest order given by (\ref{elements}). The
advantage of having a theory in the form of infinite series consists
in the possibility to calculate the determinant of the above matrix
perturbatively in $1/m^2$. Assuming that the lower
right subdeterminant  of the matrix is not zero (it is not zero in
the flat case) one has
\be
\det \hat\Phi = (A-BD^{-1}C ) \det D  \; , \qquad
\det D \neq 0
\ee
Converting the matrix D perturbatively
\be
D^{-1} =  - \frac{1}{m^2 G^{00}} \delta_i^j +
O\left(\frac{1}{m^4}\right)
\ee
we get
\be
A-BD^{-1}C = R G^{00} 2(a_1+a_2) + R^{00} (2a_4+2a_5-1) +
O\left(\frac{1}{m^2}\right)
\ee
So consistency with the flat limit imposes at this order in $m^2$ two
conditions on the five non-minimal couplings in the lagrangian
(\ref{higher}) and we are left with a three parameters family of
theories:
\be
a_1=\frac{\xi_1}{2} , \quad
a_2=-\frac{\xi_1}{2} ,  \quad
a_3=\frac{\xi_3}{2}  ,  \quad
a_4=\frac{1}{2}-\xi_2 , \quad
a_5=\xi_2 .
\ee
The action (\ref{higher}) then takes the form:
\bea
&&
S_H =\int d^D x\sqrt{-G} \biggl\{ \frac{1}{4} \nabla_\mu H \nabla^\mu H
-\frac{1}{4} \nabla_\mu H_{\nu\rho} \nabla^\mu H^{\nu\rho}
-\frac{1}{2} \nabla^\mu H_{\mu\nu} \nabla^\nu H
+\frac{1}{2} \nabla_\mu H_{\nu\rho} \nabla^\rho H^{\nu\mu}
\nonumber
\\&&
\qquad
{}+\frac{\xi_1}{4} R H_{\alpha\beta} H^{\alpha\beta}
-\frac{\xi_1}{4} R H^2
+\frac{1-2\xi_2}{4} R^{\alpha\beta} H_{\alpha\sigma} H_\beta{}^\sigma
+\frac{\xi_2}{2} R^{\alpha\beta} H_{\alpha\beta} H
\nonumber\\&&\qquad{}
+\frac{\xi_3}{2} R^{\mu\alpha\nu\beta} H_{\mu\nu} H_{\alpha\beta}
- \frac{m^2}{4} H_{\mu\nu} H^{\mu\nu} + \frac{m^2}{4} H^2
+ O\Bigl(\frac{1}{m^2}\Bigr)
 \biggr\}
\label{consistent}
\eea

In this case the rank of the matrix $\hat\Phi$ is equal to $D-1$ and
one can construct from the conservation conditions for the secondary
constraints
\be
\n_0 \varphi^{(2)}_\nu = \n_0\n^\mu E_{\mu\nu} = \hat\Phi_\nu{}^\mu
\ddot H_{0\mu} + \ldots
\ee
one linear combination which does not contain acceleration $\ddot
H_{0\mu}$:
\bea
&&
\Bigl[G^{00} -\frac{1}{m^2}R^{00} +O\Bigl(\frac{1}{m^4}\Bigr) \Bigr]
\n_0 \varphi^{(2)}_0
+\Bigl[G^{0i} -\frac{1}{m^2}R^{0i} +O\Bigl(\frac{1}{m^4}\Bigr) \Bigr]
\n_0 \varphi^{(2)}_i
\nonumber\\
&& \qquad\qquad{}
= G^{0\nu}\n_0\n^\mu E_{\mu\nu}
- \frac{1}{m^2} R^{0\nu} \n_0\n^\mu E_{\mu\nu}
+ O\Bigl(\frac{1}{m^4}\Bigr)
\eea
After excluding the accelerations $\ddot H_{ij}$ by means of
equations of motion this expression will represent a new constraint
restricting the trace of the field $H$. To make it covariant we can
add spatial derivatives of the secondary constraints
$\sim \n_i\n^\mu E_{\mu\nu}$:
\bea
\varphi^{(3)} &\approx& \n^\mu \n^\nu E_{\mu\nu}
- \frac{1}{m^2} R^{\alpha\nu} \n_\alpha\n^\mu E_{\mu\nu}
+ O\Bigl(\frac{1}{m^4}\Bigr)
\nonumber\\
&=&
(m^2-\xi_1 R) (\n^2 H - \n^\alpha \n^\beta H_{\alpha\beta})
\nonumber\\&&{}
+\xi_2 R^{\alpha\beta} (\n^2 H_{\alpha\beta} + \n_\alpha\n_\beta H -
\n^\nu\n_\alpha H_{\beta\nu} - \n^\nu\n_\beta H_{\alpha\nu} )
\nonumber\\&&{}
+ \xi_3 R^{\mu\alpha\nu\beta} \n_\mu \n_\nu H_{\alpha\beta}
+ (1+2\xi_2+2\xi_3) \n^\mu R^{\alpha\beta}
(\n_\mu H_{\alpha\beta} - \n_\alpha H_{\mu\beta})
\nonumber\\&&{}
+ (\frac{1}{2}-2\xi_1+\xi_2) \n^\alpha R (\n_\alpha H - \n^\mu
H_{\alpha\mu})
+ (-\xi_1+\frac{\xi_2}{2}) H \n^2 R
\nonumber\\&&{}
+ H_{\alpha\beta} \Bigl[ (1+\xi_2+\xi_3) \n^2 R^{\alpha\beta}
+ (-\frac{1}{2}+\xi_1-\xi_2-\frac{\xi_3}{2}) \n^\alpha\n^\beta R
\nonumber\\&&{}
- \xi_3 R^{\alpha\mu} R^\beta{}_\mu
+ \xi_3 R^{\alpha\mu\beta\nu} R_{\mu\nu}  \Bigr]
+ O\Bigl(\frac{1}{m^2}\Bigr)
\label{phi3}
\eea

Derivatives of the field $H_{\mu\nu}$ enter this expression in such a
way that it does not contain the accelerations $\ddot H_{00}$,
$\ddot H_{0\mu}$  and the velocity $\dot H_{00}$. This velocity
does not appear in (\ref{phi3}) after excluding the accelerations
$\ddot H_{ij}$ because the equations of motion do not contain
$\dot H_{00}$ either. It means that just like in the flat case
the conservation condition $\dot\varphi^{(3)}\approx 0$ leads to
another new constraints $\varphi^{(4)}$ and the last acceleration
$\ddot H_{00}$ is defined from $\dot\varphi^{(4)}\approx 0$. The
total number of constraints coincides with that in the flat spacetime.

The constraints $\varphi^{(3)}$ and $\varphi^{(2)}_\mu$ can be solved
perturbatively in $1/m^2$ with respect to the trace and the
longitudinal part of $H_{\mu\nu}$:
\be
\varphi^{(3)} \sim H + O\Bigl(\frac{1}{m^2}\Bigr) , \qquad
\varphi^{(2)}_\mu \sim \n^\mu H_{\mu\nu} + O\Bigl(\frac{1}{m^2}\Bigr)
\label{nice}
\ee
and used for simplifying the form of the original equations of motion.
It is convenient to take these equations in the following linear
combination which does not contain the term $m^2H$:
\bea
&&E_{\mu\nu} - \frac{1}{D-1} G_{\mu\nu} G^{\alpha\beta}E_{\alpha\beta}=
\nonumber\\&&{}
= \n^2 H_{\mu\nu} + \n_\mu\n_\nu H - \n_\mu\n^\sigma H_{\sigma\nu}
- \n_\nu\n^\sigma H_{\sigma\mu}
+\frac{1}{D-1} G_{\mu\nu} (\n^\alpha\n^\beta H_{\alpha\beta} -\n^2 H)
\nonumber\\&&{}\qquad
- m^2 H_{\mu\nu} + \xi_1 RH_{\mu\nu}
+(\xi_3+2) R_\mu{}^\alpha{}_\nu{}^\beta H_{\alpha\beta}
- (\frac{1}{2}+\xi_2) (R_\mu{}^\alpha H_{\alpha\nu}
+ R_\nu{}^\alpha H_{\alpha\mu})
\nonumber\\&&{}\qquad
+ \xi_2 R_{\mu\nu} H
- \frac{\xi_2}{D-1} G_{\mu\nu}RH
+ \frac{\xi_2-\xi_3-1}{D-1} G_{\mu\nu} R^{\alpha\beta} H_{\alpha\beta}
+ O\Bigl(\frac{1}{m^2}\Bigr) {.}
\eea
Adding to these equations a suitable combination of the constraints
(\ref{nice}) we obtain that in the lowest order in $1/m^2$ the spin 2
massive field is described by the conditions
\bea
&& \n^2 H_{\mu\nu} - m^2 H_{\mu\nu} + \xi_1 RH_{\mu\nu}
+(\xi_3+2) R_\mu{}^\alpha{}_\nu{}^\beta H_{\alpha\beta}
- (\frac{1}{2}+\xi_2) (R_\mu{}^\alpha H_{\alpha\nu}
+ R_\nu{}^\alpha H_{\alpha\mu})
\nonumber\\&&\qquad\qquad
+ \frac{\xi_2-\xi_3-1}{D-1} G_{\mu\nu} R^{\alpha\beta} H_{\alpha\beta}
+ O\Bigl(\frac{1}{m^2}\Bigr) = 0 {,}
\nonumber\\&&
H + O\Bigl(\frac{1}{m^2}\Bigr) = 0 {,} \qquad
\n^\mu H_{\mu\nu} + O\Bigl(\frac{1}{m^2}\Bigr) = 0
\label{finhigh}
\eea
and also by the $D$ primary constraints $E^0{}_\mu$.
We see that even in this lowest order in $m^2$ not all
non-minimal terms in the equations are arbitrary. Consistency with
the flat limit leaves only three arbitrary parameters while the number
of different non-minimal terms in the equations is four.

However, if gravitational field is also subject to some dynamical
equations of the form $R_{\mu\nu}=O(1/m^2)$ then the system
(\ref{finhigh}) contains only one non-minimal coupling in the lowest
order
\bea
&& \n^2 H_{\mu\nu} - m^2 H_{\mu\nu}
+(\xi_3+2) R_\mu{}^\alpha{}_\nu{}^\beta H_{\alpha\beta}
+ O\Bigl(\frac{1}{m^2}\Bigr) = 0 {,}
\nonumber\\&&
H + O\Bigl(\frac{1}{m^2}\Bigr) = 0 {,} \qquad
\n^\mu H_{\mu\nu} + O\Bigl(\frac{1}{m^2}\Bigr) = 0 {,}
\nonumber\\&&
R_{\mu\nu} + O\Bigl(\frac{1}{m^2}\Bigr) = 0
\label{ricci}
\eea
and is consistent for any its value.

Requirement of causality does not impose any restrictions on the
couplings in this order. The characteristic matrix of (\ref{finhigh})
is non-degenerate, second derivatives enter in the same way as in the
flat spacetime, and hence the light cones of the field $H_{\mu\nu}$
described by (\ref{finhigh}) are the same as in the flat case.
Propagation is causal for any values of $\xi_1$, $\xi_2$, $\xi_3$. In
higher orders in $1/m^2$ situation becomes more complicated and we
expect that requirement of causality may give additional restrictions
on the non-minimal couplings.

Concluding this section we would like to stress once more that
the theory (\ref{consistent}) admits any gravitational background and
so no inconsistencies arise if one treats gravity as dynamical field
satisfying Einstein equations with the energy - momentum tensor for
the field $H_{\mu\nu}$. The action for the system of interacting
gravitational field and massive spin 2 field and the Einstein
equations for it are:
\begin{eqnarray}
&&S = S_E + S_H {,} \qquad
S_E = - \frac{1}{\kappa^{D-2}} \int d^D x \sqrt{-G} R {,}
\nonumber\\&&
R_{\mu\nu} - \frac{1}{2} G_{\mu\nu} R = \kappa^{D-2} T^H_{\mu\nu} {,}
\qquad
T^H_{\mu\nu} = \frac{1}{\sqrt{-G}}\frac{\delta S_H}{\delta G^{\mu\nu}}
\end{eqnarray}
with $S_H$ given by (\ref{consistent}). However, making the metric
dynamical we change the structure of the second derivatives by
means of nonminimal terms $\sim RHH$ which  can spoil causal
propagation of both metric and massive spin 2 field \cite{aragone}.
This will impose extra restrictions on the parameters of the
theory. Also, one can consider additional requirements the
theory should fulfill, e.g. tree level unitarity of graviton -
massive spin 2 field interaction \cite{cpd}.

\section{Open string theory in background of massive spin 2 field}

In this section we will consider sigma-model description of an open
string interacting with two background fields -- massless graviton
$G_{\mu\nu}$ and second rank symmetric tensor field $H_{\mu\nu}$ from
the first massive level of the open string spectrum. We will
show that effective equations of motion for these fields are of the
form (\ref{ricci}) and explicitly calculate the coefficient $\xi_3$ in
these equations in the lowest order in $\alpha'$.

Classical action has the form
\begin{equation}
S=S_0+S_I=
\frac{1}{4\pi\alpha'}\int_M \!\!d^2z\sqrt{g}
      g^{ab}\partial_ax^\mu\partial_bx^\nu G_{\mu\nu}
+\frac{1}{2\pi\alpha'\mu}\int_{\partial M} e dt \;
      H_{\mu\nu}\dot{x}^\mu\dot{x}^\nu
\label{actstring}
\end{equation}
Here $\mu,\nu=0,\ldots,D-1$; $a,b=0,1$ and we introduced the notation
$\dot{x}^\mu=\frac{dx^\mu}{edt}$. The first term $S_0$ is an integral
over two-dimensional string world sheet $M$ with metric $g_{ab}$ and
the second $S_I$ represents a one-dimensional integral over its
boundary with einbein $e$. We work in euclidian signature and
restrict ourselves to flat world sheets with straight boundaries. It
means that both two-dimensional scalar curvature and extrinsic
curvature of the world sheet boundary vanish and we can always choose
such coordinates that $g_{ab}=\delta_{ab}$, $e=1$.

Theory has two dimensionful parameters. $\alpha'$ is fundamental
string length squared, $D$-dimensional coordinates $x^\mu$ have
dimension $\sqrt{\alpha'}$. Another parameter $\mu$ carries dimension
of inverse length in two-dimensional field theory (\ref{actstring})
and plays the role of renormalization scale. It is introduced in
(\ref{actstring}) to make the background field $H_{\mu\nu}$
dimensionless. In fact, power of $\mu$ is responsible for the number
of massive level to which a background field belongs because one
expects that open string interacts with a field from $n$-th massive
level through the term
$$
\mu^{-n} (\alpha')^{-\frac{n+1}{2}}
\int_{\partial M} e dt\; \dot{x}{}^{\mu_1} \ldots
\dot{x}{}^{\mu_{n+1}} H_{\mu_1 \ldots \mu_{n+1}} (x)
$$

The action (\ref{actstring}) is non-renormalizable from the point of
view of two-dimensional quantum field theory. Inclusion of
interaction with any massive background produces in each loop an
infinite number of divergencies and so requires an infinite number of
different massive fields in the action. But massive modes from the
$n-$th massive level give vertices proportional to $\mu^{-n}$ and so
they cannot contribute to renormalization of fields from lower
levels. Of course, this argument assumes that we treat the theory
perturbatively defining propagator for $X^\mu$ only by the term with
graviton in (\ref{actstring}). Now we will use such a scheme to
carry out renormalization of (\ref{actstring}) dropping all the terms
$O(\mu^{-2})$.

Varying (\ref{actstring}) one gets classical equations of motion with
boundary conditions:
\begin{eqnarray}
&& g^{ab} D_a \partial_b x^\alpha \equiv
g^{ab}(\partial_a \partial_b x^\alpha + \Gamma^\alpha_{\mu\nu}(G)
\partial_a x^\mu \partial_b x^\nu) = 0,
\nonumber\\&&
\left. G_{\mu\nu} \partial_n x^\mu \right|_{\partial M} -
\frac{2}{\mu} {\cal D}^2_t x^\mu H_{\mu\nu} +\frac{1}{\mu}
{\dot x}^\mu {\dot x}^\lambda (\nabla_\nu H_{\mu\lambda}
-\nabla_\mu H_{\nu\lambda} - \nabla_\lambda H_{\mu\nu}) = 0
\label{class}
\end{eqnarray}
where  $\partial_n=n^a \partial_a$, $n^a$ is unit inward normal
vector to the world sheet boundary and ${\cal D}^2_t x^\mu = {\ddot
x}^\mu+\Gamma^\mu_{\nu\lambda}(G) {\dot x}^\nu {\dot x}^\lambda$ is
the derivative covariant with respect to D-dimensional
diffeomorphisms.

To calculate divergencies of one-loop effective action
$\Gamma^{(1)}_{div}$ we need to expand (\ref{actstring}) up to
the second order in normal coordinates
\begin{equation}
x^\mu=\bar{x}^\mu+\xi^\mu - \frac{1}{2} \Gamma^\mu_{\alpha\beta}
\xi^\alpha \xi^\beta +O(\xi^3)
\label{normal}
\end{equation}
around a solution ${\bar x}^\mu$ of classical equations of motion
(\ref{class}):
\begin{eqnarray}
&&\Gamma^{(1)}= \frac{1}{2}\mbox{Tr}\ln
\left( S_{0\alpha\beta}+\frac{1}{2\pi\alpha'\mu}V_{\alpha\beta}
  \right)
\nonumber\\&&{}
  =\frac{1}{2}\mbox{Tr}\ln S_{0\alpha\beta}
  +\frac{1}{2\mu}\mbox{Tr}V_{\gamma\alpha}G_{0}^{\gamma\beta}
  +O(\mu^{-2})
\label{trln}
\\&&\qquad\qquad
 S_{0\alpha\beta}=\frac{\delta^2 S_0}{\delta\xi^\alpha\delta\xi^\beta};\qquad
  \frac{1}{2\pi\alpha'\mu}V_{\alpha\beta}=
     \frac{\delta^2S_I}{\delta\xi^\alpha\delta\xi^\beta}
\nonumber
\end{eqnarray}
Here Green function $G_0^{\gamma\beta}$ of the operator
$S_{0\alpha\gamma}$ is defined as
$$
2\pi\alpha' S_{0\alpha\gamma}G_0^{\gamma\beta}=\delta^\beta_\alpha
$$
The terms $O(\mu^{-2})$ in (\ref{trln})  contribute to
renormalization of the second and higher massive levels only and will
be omitted from here on.

Explicit calculations give the following expressions for
$V_{\alpha\beta}$ and $S_{0\alpha\beta}$
\begin{eqnarray}
S_{0\alpha\beta}(z,z')&=&-\frac{1}{2\pi\alpha'}\left(
  G_{\alpha\beta}g^{ab}D_aD_b
  +g^{ab}\partial_ax^\mu\partial_bx^\nu R_{\alpha\mu\beta\nu}
  \right)\delta(z,z')-
\nonumber\\
&&{}-\frac{1}{2\pi\alpha'}\delta_{\partial M}(z)
       G_{\alpha\beta}n^aD_{a}\delta(z,z')
\nonumber\\
V_{\alpha\beta}(z,z')&=&\delta_{\partial M}(z)
 \sum_{k=0}^{2}V^{(k)}_{\alpha\beta}(z)
 ({\cal D}_t)^k  \delta(z,z')
\nonumber\\
V^{(0)}_{\alpha\beta}&=&
  -2 \nabla_\beta H_{\mu\alpha}{\cal D}^2_t x^\mu
+ \dot{x}^\mu\dot{x}^\nu \left( \nabla_\alpha \nabla_\beta H_{\mu\nu}
- 2 \nabla_\beta \nabla_\nu H_{\mu\alpha}
-2 H_{\alpha\lambda}R^\lambda{}_{\mu\beta\nu}
\right)
\nonumber\\
 V^{(1)}_{\alpha\beta}&=&
    2 \dot{x}^\mu \left( \nabla_\alpha H_{\mu\beta} -
     \nabla_\beta H_{\mu\alpha} - \nabla_\mu H_{\alpha\beta}   \right)
\nonumber\\
  V^{(2)}_{\alpha\beta}&=&-2 H_{\alpha\beta}
\end{eqnarray}

Here delta-function of the boundary $\delta_{\partial M}(z)$ is
defined as
\begin{eqnarray}
&& \int_M\delta_{\partial M}(z)
V(z)\sqrt{g(z)} d^2z =\int_{\partial M}V|_{z\in\partial M} e(t) dt
\end{eqnarray}

Using dimensional regularization and following the procedure
\cite{mcavity} one gets divergences of the Green function and its
derivatives:
\begin{eqnarray}
  \left.\mbox{Tr\ }\ln S_{0;\alpha\beta}\right|^{div}&=&
    \frac{\mu^{-\varepsilon}}{2\pi\varepsilon}
    \int_{M} d^{2+\varepsilon} z\sqrt{g}
    g^{ab}\partial_ax^\mu\partial_bx^\nu R_{\mu\nu}
\nonumber\\
  \left.\phantom{\frac{D}{edt}}G^\alpha_{0\beta}\right|_{\partial M}^{div}
  &=&{}-\frac{\mu^{-\varepsilon}}{\pi\varepsilon}\delta^\alpha_\beta
\\
  \left. {\cal D}_t G^\alpha_{0\beta}\right|_{\partial M}^{div}&=&0
\\
  \left. {\cal D}^2_t G^\alpha_{0\beta}\right|_{\partial M}^{div}
  &=&{}-\frac{\mu^{-\varepsilon}}{2\pi\varepsilon}
\left.  g^{ab}\partial_ax^\mu\partial_bx^\nu R^\alpha{}_{\mu\nu\beta}
\right|_{\partial M}
\label{div}
\end{eqnarray}

As a result, divergent part of the one loop effective
action has the form
\begin{eqnarray}
\Gamma^{(1)}_{div}&=&\frac{\mu^{-\varepsilon}}{4\pi\varepsilon}
    \int_{M} d^{2+\varepsilon} z\sqrt{g}
    g^{ab}\partial_ax^\mu\partial_bx^\nu R_{\mu\nu}
\nonumber\\&&{}
    -\frac{\mu^{-\varepsilon-1}}{2\pi\varepsilon}\int_{\partial M}
    dt e(t) \dot{x}^\mu\dot{x}^\nu
   \left( \nabla^2 H_{\mu\nu} - 2R_\mu{}^\alpha H_{\alpha\nu}
+ R_\mu{}^\alpha{}_\nu{}^\beta H_{\alpha\beta} \right)
\end{eqnarray}
which leads to one-loop renormalization of the
background fields:
\begin{eqnarray}
\stackrel{\circ}{G}_{\mu\nu}&=&\mu^\varepsilon
  G_{\mu\nu}-\frac{\alpha'\mu^\varepsilon}{\varepsilon}R_{\mu\nu}
\nonumber\\
\stackrel{\circ}{H}_{\mu\nu}&=&\mu^\varepsilon H_{\mu\nu}
  +\frac{\alpha'\mu^\varepsilon}{\varepsilon}
  \left( \nabla^2 H_{\mu\nu} -2 R^\sigma{}_{(\mu} H_{\nu)\sigma}
  + R_\mu{}^\alpha{}_\nu{}^\beta H_{\alpha\beta} \right)
\label{1loop}
\end{eqnarray}
with circles denoting bare values of the fields. We would like to
stress once more that higher massive levels do not influence the
renormalization of any given field from the lower massive levels and
so the result (\ref{1loop}) represents the full answer for
perturbative one-loop renormalization of $G_{\mu\nu}$ and
$H_{\mu\nu}$.

Now to impose the condition of Weyl invariance of the theory at the
quantum level we should calculate the trace of energy momentum tensor
in $d=2+\varepsilon$ dimension:
\begin{equation}
  T(z) = g_{ab}(z) \frac{\delta S}{\delta g_{ab}(z)} =
\frac{\varepsilon\mu^{-\varepsilon}}{8\pi\alpha'}
g^{ab}(z)\partial_ax^\mu\partial_bx^\nu G_{\mu\nu}
-\frac{\mu^{-1-\varepsilon}}{4\pi\alpha'} H_{\mu\nu} \dot{x}^\mu
\dot{x}^\nu \delta_{\partial M}(z)
\end{equation}
performing one-loop renormalization of the composite operators
$g^{ab}(z)\partial_a x^\mu\partial_bx^\nu G_{\mu\nu}$ and
$H_{\mu\nu} \dot{x}^\mu \dot{x}^\nu$.

Divergencies in vacuum expectations values of the operators
\begin{equation}
<H_{\mu\nu} \dot{x}^\mu \dot{x}^\nu > =
\frac{\int {\cal D}x \, e^{-S[x]} H_{\mu\nu} \dot{x}^\mu \dot{x}^\nu}%
{\int {\cal D}x \, e^{-S[x]} }
\label{expect}
\end{equation}
are calculated by standard background field method with the use of
normal coordinates (\ref{normal}) as quantum fields.

In the one loop approximation it is enough to expand the
composite operator and the action in (\ref{expect}) up the to second
order in normal coordinates and we neglect the terms more than
linear in $H_{\mu\nu}$ as giving only contributions to
renormalization of higher massive levels:
\begin{eqnarray}
<\dot{x}^\mu \dot{x}^\nu H_{\mu\nu}  > &=&
\dot{\bar x}^\mu \dot{\bar x}^\nu H_{\mu\nu} (\bar x)
+ <\xi^\alpha \xi^\beta>_0 \dot{\bar x}^\mu \dot{\bar x}^\nu
 \biggl(\frac{1}{2}  \nabla_\alpha \nabla_\beta H_{\mu\nu} (\bar x) +
      R^\sigma{}_{\alpha\beta\mu} H_{\sigma\nu} (\bar x) \biggr)
\nonumber\\&&{}
+ < {\cal D}_t \xi^\mu {\cal D}_t \xi^\nu>_0 H_{\mu\nu} (\bar x)
+ < {\cal D}_t \xi^\nu \xi^\alpha>_0 2 \dot{\bar x}^\mu \nabla_\alpha
    H_{\mu\nu} (\bar x)
\end{eqnarray}
Here
\begin{eqnarray}
<\xi^\alpha (z) \xi^\beta (z')> \equiv
\frac{\int {\cal D} \xi \, e^{- \frac{1}{2} S_{0;\mu\nu} [\bar x]
\xi^\mu \xi^\nu} \xi^\alpha (z) \xi^\beta (z')  }%
{\int {\cal D} \xi \, e^{- \frac{1}{2} S_{0;\mu\nu} [\bar x]
\xi^\mu \xi^\nu}}
 = 2\pi G_0^{\alpha\beta} (z,z')
\end{eqnarray}
is Green function for the fields $\xi^\mu$.

Using (\ref{div}) one gets:
\begin{eqnarray}
<\dot{x}^\mu \dot{x}^\nu H_{\mu\nu}  > &=&
\mu^{-\varepsilon}\dot{\bar x}^\mu \dot{\bar x}^\nu
\biggl( H_{\mu\nu}  + \frac{2}{\varepsilon} R^\alpha{}_\mu
H_{\alpha\nu} - \frac{1}{\varepsilon} \nabla^2 H_{\mu\nu} \biggr)
\nonumber\\&&{}
- \frac{\mu^{-\varepsilon}}{\varepsilon}  g^{ab} \partial_a \bar x^\mu
\partial_b \bar x^\nu R_\mu{}^\alpha{}_\nu{}^\beta H_{\alpha\beta}
+ (fin)
\end{eqnarray}
Renormalized composite operator should have finite expectation value
and so the renormalization should have the form (we use minimal
subtraction scheme):
\begin{eqnarray}
\biggl( \dot{x}^\mu \dot{x}^\nu H_{\mu\nu} \biggr)_0 &=&
\mu^{-\varepsilon}
\biggl[ \dot x^\mu \dot x^\nu \Bigl( H_{\mu\nu}  +
\frac{2}{\varepsilon} R^\alpha{}_\mu H_{\alpha\nu} -
\frac{1}{\varepsilon} \nabla^2 H_{\mu\nu} \bigr) \biggr]
\nonumber\\&&{}
- \frac{\mu^{-\varepsilon}}{\varepsilon}
\biggl[ \dot x^\mu \dot x^\nu R_\mu{}^\alpha{}_\nu{}^\beta
H_{\alpha\beta} \biggr] + (fin)
\end{eqnarray}
where $(\ldots)_0$ and $[\ldots]$ stands for bare and
renormalized operators respectively. Expressing bare values of
background field $\stackrel{\circ}{H}_{\mu\nu}$ in terms of the
renormalized ones (\ref{1loop}) we finally see that in the lowest
order in $\alpha'$ the operator does not receive any renormalization:
\begin{equation}
\bigl( \dot{x}^\mu \dot{x}^\nu \stackrel{\circ}{H}_{\mu\nu} \bigr)_0 =
\mu^{-\varepsilon}
\bigl[ \dot x^\mu \dot x^\nu  H_{\mu\nu} \bigr]
\end{equation}

After the same but more tedious calculations we get the following
renormalization of another composite operator:
\begin{eqnarray}&&
(g^{ab}\partial_a x^\mu \partial_b x^\nu
   \stackrel{\circ}{G}_{\mu\nu})_0 =\mu^\varepsilon
   \Bigl[g^{ab}\partial_a x^\mu \partial_b x^\nu (G_{\mu\nu}
   -\frac{\alpha'}{\varepsilon} R_{\mu\nu})\Bigr]
\\&&\qquad\qquad{}
+\frac{\alpha'\mu^{-1+\varepsilon}}{\varepsilon}
    \left[H_{\alpha}{}^\alpha\delta_{\partial M}''(z)
   + {\cal D}^2_tx^\mu (\nabla_\mu H_\alpha{}^\alpha
   - 4 \nabla^\alpha H_{\alpha\mu}) \delta_{\partial M}(z)
\right.
\nonumber\\&&\qquad\qquad{}
\left.
   + {\dot x}^\mu {\dot x}^\nu (
    \nabla_\mu \nabla_\nu H_\alpha{}^\alpha
   - 4 \nabla^\alpha \nabla_{(\mu} H_{\nu)\alpha}
   + 2 \nabla^2 H_{\mu\nu} - 2 R_\mu{}^\alpha{}_\nu{}^\beta
     H_{\alpha\beta} ) \delta_{\partial M}(z) \right]
\nonumber
\end{eqnarray}
and build the renormalized operator of the energy momentum tensor
trace:
\begin{eqnarray}
8\pi[T]&=&{}
- \Bigl[ g^{ab} \partial_a x^\mu \partial_b x^\nu
E_{\mu\nu}^{(0)}(x)\Bigr] + \frac{2}{\mu} \delta_{\partial M}(z)
   \Bigl[ {\dot x}^\mu {\dot x}^\nu E^{(1)}_{\mu\nu}(x) \Bigr]
\nonumber\\&&{}
+ \frac{1}{\mu} \delta_{\partial M}(z)
   \Bigl[ {\cal D}^2_t x^\mu E^{(2)}_\mu(x) \Bigr]
+ \frac{1}{\mu} \delta''_{\partial M}(z) \Bigl[ E^{(3)}(x) \Bigr]
\end{eqnarray}
where
\begin{eqnarray}
E^{(0)}_{\mu\nu} (x)&=& R_{\mu\nu} + O(\alpha')
\nonumber\\
E^{(1)}_{\mu\nu}(x)&=&
\nabla^2 H_{\mu\nu} - \nabla^\alpha \nabla_\mu H_{\alpha\nu}
- \nabla^\alpha \nabla_\nu H_{\alpha\mu}
\nonumber\\&&{}
- R_\mu{}^\alpha{}_\nu{}^\beta H_{\alpha\beta}
+ \frac{1}{2} \nabla_\mu \nabla_\nu H_\alpha{}^\alpha
- \frac{1}{\alpha'} H_{\mu\nu}  + O(\alpha')
\nonumber\\
E^{(2)}_\mu(x)&=& \nabla_\mu H_\alpha{}^\alpha
- 4 \nabla^\alpha H_{\alpha\mu} + O(\alpha')
\nonumber\\
E^{(3)}(x)&=&H_\alpha{}^\alpha + O(\alpha')
\label{E}
\end{eqnarray}
Terms of order $O(\alpha')$ arise from the higher loops contributions.

The requirement of quantum Weyl invariance tells that all $E(x)$
in (\ref{E}) should vanish and so they are interpreted as effective
equations of motion for background fields. They contain vacuum
Einstein equation for graviton (in the lowest order in $\alpha'$),
curved spacetime generalization of the mass shell condition for the
field $H_{\mu\nu}$ with the mass $m^2=(\alpha')^{-1}$ and $D+1$
additional constraints on the values of this fields and its first
derivatives. Taking into account these constraints and the Einstein
equation we can write our final equations arising from the Weyl
invariance of string theory in the form:
\begin{eqnarray}
&& \nabla^2 H_{\mu\nu}
+ R_\mu{}^\alpha{}_\nu{}^\beta H_{\alpha\beta}
- \frac{1}{\alpha'} H_{\mu\nu} + O(\alpha') = 0 {,}
\nonumber\\&&
\nabla^\alpha H_{\alpha\nu} + O(\alpha') =0 {,} \qquad
H^\mu{}_\mu +O(\alpha') =0 {,}
\nonumber\\&&
R_{\mu\nu} +O(\alpha')= 0 {.}
\label{final}
\end{eqnarray}
They coincide with the equations found in the previous section
(\ref{ricci}) with the value of non-minimal coupling $\xi_3=-1$.

In fact, Einstein equations should not be vacuum ones
but contain dependence on the field $H_{\mu\nu}$ through its energy -
momentum tensor $T^H_{\mu\nu}$. Our calculations could not produce
this dependence because such dependence is expected to arise only if
one takes into account string world sheets with non-trivial topology
and renormalizes new divergencies arising from string loops
contribution \cite{fs,clny}.
\be
R_{\mu\nu} +O(\alpha')= T^H_{\mu\nu} - \frac{1}{D-2}
 T^H{}^\alpha{}_\alpha {,}
\ee
where explicit form of the lowest contributions to the
energy-momentum tensor $T^H_{\mu\nu}$ can be determined
only from sigma model on world sheets with topology of annulus.

In order to determine whether the equations (\ref{final}) can be
deduced from an effective lagrangian (and to find this lagrangian)
one would need two-loop calculations in the string sigma-model.
Two-loop contributions to the Weyl anomaly coefficients $E^{(i)}$  are
necessary because the effective equations of motion
(\ref{E},\ref{final}) are not the equations directly following from a
lagrangian but some combinations of them similar to (\ref{finhigh}).
In order to reverse the procedure of passing from the original
lagrangian equations to (\ref{finhigh}) one would need the next to
leading contributions in the conditions for $\nabla^\mu H_{\mu\nu}$
and $H_{\mu\nu}$ (\ref{final}).

\section{Conclusion}

Let us summarize the obtained results.  We investigated the problem
of consistency of the equations of motion for spin-2 massive field in
curved spacetime and found that two different description
of this field are possible. First, for gravitational
background satisfying vacuum Einstein equations (\ref{einst}) one can
build an action leading to consistent equations including the
tracelessness and transversality conditions. Of course, such
gravitational backgrounds include all popular vacuum solutions of
Einstein equations such as constant curvature spacetimes,
Schwarzschild solution,  plane waves etc. It would be
interesting to investigate properties of the massive spin 2 field
dynamics on these specific exact solutions.

Another possibility (naturally arising in string theory) consists in
building the theory as perturbation series in inverse mass. In the
lowest order no problems of consistency with the flat space limit and
causality arise and equations of motion have the form
(\ref{finhigh}).

Then we calculated the equations for the massive
spin-2 background field arising in sigma model approach to string
theory from the condition of quantum Weyl invariance in the lowest
order in $\alpha'$. The explicit form of the derived equations
(\ref{final}) appears to be a particular case of the
general equations in field theory (\ref{finhigh}).
We expect that in general in each order in $\alpha'$ the situation
remains the same and it is possible to construct the part of string
effective action quadratic in massive background field which should
lead to generalized mass-shell, tracelessness and transversality
conditions.

To determine this part of the bosonic string effective action
completely one should also consider other massless background field
including the dilaton $\phi(x)$ and antisymmetric tensor
$B_{\mu\nu}(x)$.  Inclusion of dilaton will require investigation of
strings with curved world sheets and with non-vanishing extrinsic
curvature on the boundary which complicates the sigma-model
calculations. Interaction with massless antisymmetric tensor will be
especially interesting in presence of a D-brane because such a
system is a source of non-commutative geometry in string theory
(see \cite{SW} and references therein). In the limit when components
of $B_{\mu\nu}$ along the brane are large there should arise some
non-commutative counterpart of the spin two massive field theory and
we hope to derive its explicit form in the future.

Also it would be interesting to repeat our analysis in the case
of closed string which contains the fourth rank tensor at the lowest
massive level. From the field theoretical point of view
investigation of such higher spin fields interacting with gravity
will require to generalize analysis made in the Section~2
to the case of arbitrary spin fields whose
dynamics is governed by more complex lagrangians with auxiliary
fields \cite{singh}. We leave this investigation for the future work.

\section*{Acknowledgements}

I.L.B. is grateful to S.~Kuzenko, H.~Osborn, B.~Ovrut, A.~Tseytlin
and G.~Veneziano for useful discussions of some aspects of this work.
The work of I.L.B., V.A.K. and V.D.P. was supported by GRACENAS
grant, project 97-6.2-34 and RFBR grant, project 99-02-16617;
the work of I.L.B. and V.D.P. was supported by RFBR-DFG grant,
project 99-02-04022 and INTAS grant N 99 0590. I.L.B. is grateful
to FAPESP and D.M.G. is grateful to CNPq for support of the research.

\end{document}